\documentclass[aps,prl,superscriptaddress,twocolumn,longbibliography]{revtex4-2}
\usepackage{bbm}
\usepackage{graphicx}
\usepackage{dcolumn}
\usepackage{bm}
\usepackage{subfigure}
\usepackage{amsmath}
\usepackage{amsfonts}
\usepackage{feynmf}
\usepackage{hyperref}

\usepackage{attachfile}

\newcommand{\bk}{\boldsymbol k}

\newcommand{\br}{\boldsymbol r}

\newcommand{\zb}{\color {black}}

\usepackage{times}

\begin{document}
\title{Layer-Locked Chiral Topological Superconductivity}

\author{Yijie Mo}
\affiliation{Guangdong Provincial Key Laboratory of Magnetoelectric Physics and Devices, State Key Laboratory of Optoelectronic Materials and Technologies, School of Physics, Sun Yat-sen University, Guangzhou 510275, China}

\author{Zhongbo Yan}
\email{yanzhb5@mail.sysu.edu.cn}
\affiliation{Guangdong Provincial Key Laboratory of Magnetoelectric Physics and Devices, State Key Laboratory of Optoelectronic Materials and Technologies, School of Physics, Sun Yat-sen University, Guangzhou 510275, China}

\date{\today}

\begin{abstract}
We uncover a universal mechanism for realizing layer-locked topological phases. Guided by it, we investigate 
the realization of layer-locked chiral topological superconductivity---the superconducting analogue of the 
quantum anomalous layer Hall effect---in a nonsymmorphic bilayer antiferromagnetic system with 
$s$-wave pairing. We identify three distinct gate-tunable topological phases and establish a direct 
correspondence between the nearly quantized layer-resolved Chern numbers and the layer-locking behavior 
of chiral Majorana edge states, vortex-core Majorana zero modes, and nearly quantized thermal Hall responses.
\end{abstract}

\maketitle

Chiral topological superconductors {\zb (CTSCs)} have attracted considerable attention 
over the past two decades owing to their novel physical properties and broad application prospects~\cite{Qi2011TISC,Alicea2012New,Leijnse2012Intro,Tanaka2012Symmetry,Stanescu2013Majorana,Beenakker2013Search,Elliott2015Colloquium,Sato2016Majorana,Kallin2016Chiral}. This class of two-dimensional topological phases features a fully gapped bulk, with gapless chiral Majorana edge spectra traversing the gap~\cite{Read2000Paired,Qi2010Chiral}. These systems do not require any symmetry protection, and {\zb their topological nature is} captured by a nonzero Chern number---the integration of Berry curvature over the Brillouin zone~\cite{Schnyder2008Classification,Kitaev2009Periodic,Ryu2010topo}. {\zb Beyond chiral Majorana edge states (CMESs), CTSCs} exhibit two other defining features. First, vortices in systems with odd Chern numbers host robust Majorana zero modes (MZMs)~\cite{Volovik1999fermion,Read2000Paired,Ivanov2001Non}. Second, temperature gradients give rise to {\zb a Berry-curvature-driven quantized thermal Hall effect} in the low-temperature limit~\cite{Nomura2012Cross,Shimizu2015Quantum,Mo2025Coexistence}. These properties render CTSCs promising platforms for fault-tolerant topological quantum computation and thermoelectric {\zb applications}~\cite{Nayak2008Non,Sarma2015,Karzig2017Scalable,Marra2022Majorana,Dumitr1escu2012Topological,Ngampruetikorn2020Impurity}. Since 
intrinsic CTSCs (e.g., $p \pm ip$ and $d \pm id$-wave superconductors) are rare in nature, alternative approaches {\zb based on 
conventional superconductivity} have been actively pursued~\cite{Black2012Edge,Liu2013d+id,Qin2019CTSC}. A representative theoretical proposal involves utilizing the proximity effect between conventional $s$-wave superconductors and either nonmagnetic spin-orbit-coupled materials placed in a Zeeman field~\cite{Fu2008Superconducting,Fujimoto2008Topological,Zhang2008px+ip,Sato2009Non,Sau2010Generic,Alicea2010Majorana,Li2016shiba} or magnetic topological insulators~\cite{Wang2015CTSC,Tokura2019MTI} to {\zb effectively} realize CTSCs. This proposal has greatly stimulated the related experimental progress~\cite{Menard2017Two,Alexandra2019Atomic,Kezilebieke2020Topological,Li2024Observation}.

CTSCs are also known as the superconducting counterpart of the Chern insulators, as they 
share the same topological characterization and exhibit strong similarities in their bulk-boundary correspondence 
and in  Berry-curvature-driven physics.  Since both require a nonzero Chern number, their realization 
necessitates the absence of any symmetry that would force the Chern number to vanish, including time-reversal symmetry ($\mathcal{T}$), 
vertical mirror symmetry (e.g., $\mathcal{M}_{x,y}$) and space-time symmetry ($\mathcal{PT}$). 
In layered systems, however, $\mathcal{PT}$ symmetry differs from the other two. In particular, 
while it enforces a zero total Chern number by requiring opposite layer-resolved Chern numbers 
for the two $\mathcal{P}$-related layers, it does not force the Chern number of each individual 
layer to vanish. Such nonzero layer-resolved Chern numbers have recently attracted widespread interest~\cite{Gao2021layer,Chen2024Layer,Dai2022Quantum,Peng2023Intrinsic,XU2024Layer,Li2024Dissipationless,
Zhang2023Layer,Zhang2024Layer,Feng2023Layer,Liu20242024QALHE,Tian2024QALHE,Gao2024LHE,Liu2024Engineering,Wang2025Layer,Wang2025LHE,Qin2026Layer,
Tao2024Layer,Yi2024Disorder,Han2025Layer,Hu2026LHE}, as they are believed to give rise to an unconventional Hall response---the layer Hall effect.
Such an effect has been experimentally observed
in $\mathcal{PT}$-symmetric even-layer MnBi$_{2}$Te$_{4}$~\cite{Gao2021layer} via the anomalous 
Hall effect induced by a perpendicular electric field.
As with any Hall effect, its quantized counterpart---the quantum anomalous layer 
Hall effect (QALHE)~\cite{Dai2022Quantum,Liu20242024QALHE,Tian2024QALHE}---is of particular interest, as it represents 
a new type of topological phase. 
However, realizing QALHE faces intrinsic difficulties.
Phenomenologically, a quantized layer Hall effect requires that chiral edge states be strictly confined to a single layer. Such 
a condition is generally impossible to satisfy, as finite interlayer coupling inevitably allows electron 
hopping between layers. Apart from the trivial decoupled-layer limit, the only known mechanism to date is 
disorder-induced Anderson localization~\cite{Dai2022Quantum}, which relies on strong disorder to suppress interlayer hopping.
Yet disorder inevitably degrades material quality, 
and may introduce further complications that impede experimental observation.
This underscores the pressing need for an alternative mechanism that 
functions in clean systems and accommodates arbitrary interlayer coupling.

In this work, we uncover a universal, symmetry-based mechanism for realizing layer-locked 
topological phases that operates regardless of interlayer coupling strength and does not rely on disorder.  
The mechanism is surprisingly simple: when the lattices of two adjacent layers can be mutually mapped by certain 
nonsymmorphic symmetries, the interlayer hopping amplitude is symmetry-enforced to vanish along 
specific high-symmetry lines in the Brillouin zone. As a consequence, when topological phase 
transitions occur at momenta residing on these lines, the layer-resolved Berry curvatures become 
sharply concentrated near those momenta and exhibit near mutual independence. This leads to nearly 
quantized layer-resolved Chern numbers and the emergence of layer-locked chiral edge states. Given 
the evident universality of this mechanism across both insulating and superconducting platforms, we 
illustrate its utility through a superconducting system. This not only broadens the landscape of 
layer-locked phenomena but also provides a new pathway for realizing highly tunable CTSCs.

\begin{figure}[t]
	\centering
	\includegraphics[width=0.47\textwidth]{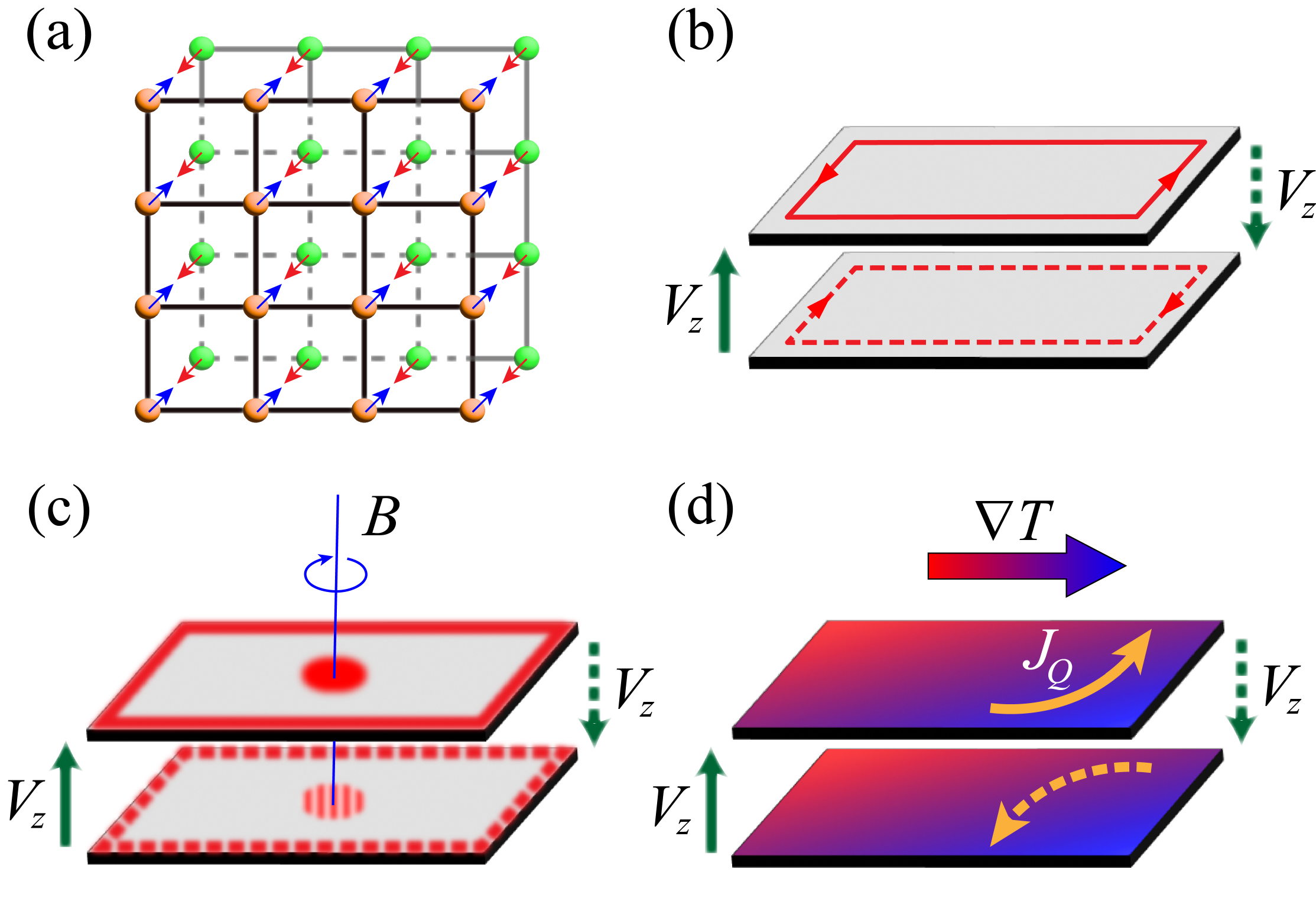}
	\caption{Schematics of the (a) lattice structure and antiferromagnetic order, (b) layer-locked CMESs, (c) layer-locked vortex-core MZMs, and (d) layer-locked thermal Hall effect.  In (b)–(d), the dashed patterns indicate a switch of the locking 
layer upon reversal of the gate voltage.
	}\label{fig1}
\end{figure}

{\it Model Hamiltonian.---}We consider a minimal bilayer system, as
sketched in Fig.~\ref{fig1}(a). The two layers are displaced by a half-unit-cell translation 
$\mathbf{t}_{d}=(a/2, a/2)$, where $a$ is the in-plane lattice constant (hereafter set to unity). 
Focusing on the lattice only, the two layers can be mutually mapped by three nonsymmorphic symmetries, 
a glide symmetry $[(\mathcal{M}_{z}|\mathbf{t}_{d})]$, and two screw symmetries $[(C_{2x}|\mathbf{t}_{x}), (C_{2y}|\mathbf{t}_{y})]$. 
Here $\mathcal{M}_{z}$ denotes mirror reflection about the midplane, $C_{2a}$ denotes $\pi$-rotation about the $a$ axis, and  
$\mathbf{t}_{x}=(1/2, 0)$ and $\mathbf{t}_{y}=(0, 1/2)$. These three symmetries dictate the interlayer 
hopping possesses a form factor $[\cos (k_{x}/2)\cos (k_{y}/2)]$ which vanishes at the Brillouin-zone boundary~\cite{Young2015DSM}.
This behavior can also be understood as a direct consequence of the lattice geometry: each site in one layer couples to four nearest neighbors in the other, resulting in four interfering hopping paths. At the zone boundary, where
$k_{x}=\pi$ or $k_{y}=\pi$, these paths interfere destructively, effectively decoupling 
the two layers regardless of the interlayer coupling strength.

We consider that the bilayer system describes a 
$\mathcal{PT}$-symmetric antiferromagnetic metal featuring hidden spin textures that originate from the local breaking of inversion symmetry~\cite{Zhang2014hidden,Chen2022hidden}.
We assume that superconductivity develops in this system either through an intrinsic interaction-driven instability in the 
$s$-wave pairing channel or via the proximity effect from an 
$s$-wave superconducting substrate~\cite{Xu2022MS}.
The corresponding  Bogoliubov-de Gennes (BdG) Hamiltonian is given by $H=\frac{1}{2}\sum_{\bk}\Psi_{\bk}^{\dag}\mathcal{H}(\bk)\Psi_{\bk}$, where  $\Psi^{\dag}_{\bk}=(c_{t,\uparrow,\bk}^{\dag} ,c_{t,\downarrow,\bk}^{\dag},c_{b,\uparrow,\bk}^{\dag},c_{b,\downarrow,\bk}^{\dag}, c_{t,\uparrow,-\bk}, c_{t,\downarrow,-\bk},c_{b,\uparrow,-\bk},c_{b,\downarrow,-\bk})$, and 
\begin{eqnarray}
\mathcal{H}(\bk)&=&[\epsilon(\bk)-\mu]\tau_{z}+4\eta\cos\frac{k_{x}}{2}\cos\frac{k_{y}}{2}\tau_{z}\sigma_{x} \nonumber\\
&&+2\lambda_{\rm so}(\sin k_{x}\tau_{z}\sigma_{z}s_{y}-\sin k_{y}\sigma_{z}s_{x})\nonumber\\
&&+M_{z}\tau_{z}\sigma_{z}s_{z}+V_{z}\tau_{z}\sigma_{z}+\Delta\tau_{y}s_{y},\label{BdGH}
\end{eqnarray}
where $\epsilon(\bk)=-2t(\cos k_{x}+\cos k_{y})$, the Pauli matrices $\{ \tau_{i} \}$, $\{ \sigma_{i} \}$ and $\{ s_{i} \}$ act on the particle-hole, layer ($t, b$) and spin ($\uparrow, \downarrow$) degrees of freedom, respectively. For notational simplicity,
all identity matrices are omitted. In $\mathcal{H}(\bk)$
the parameter $t$ denotes the intralayer nearest-neighbor hopping amplitude, 
$\eta$ denotes the interlayer nearest-neighbor hopping amplitude, and 
$\lambda_{\rm so}$ characterizes the strength of spin-orbit coupling (opposite in 
the two layers). Additionally,  $M_{z}$ is the magnetic exchange field (with the Néel vector assumed to lie along the 
$z$ direction),  and $\Delta$ is the $s$-wave pairing amplitude. We also include a layer-staggered potential of the form 
$V_{z}\tau_{z}\sigma_{z}$, which allows us to tune between the 
$\mathcal{PT}$-symmetric ($V_{z}=0$) and 
$\mathcal{PT}$-broken ($V_{z}\neq0$) regimes. Such a term can be experimentally realized via a gate voltage.

In the absence of both the magnetic exchange field ($M_{z}=0$) and the layer-staggered potential $(V_{z}=0)$, 
the normal-state band structure of this Hamiltonian exhibits three Dirac points located at $\textbf{X}=(\pi,0)$, $\textbf{Y}=(0,\pi)$, 
and $\textbf{M}=(\pi,\pi)$~\cite{Young2015DSM}. Previous studies have shown that this band structure lays the foundation for 
a diversity of topological superconducting phases, such as an exotic gapless superconducting phase featuring the coexistence 
of Bogoliubov Fermi surfaces and CMESs~\cite{Mo2025Coexistence}, unconventional topological superconductors with multiple pairs 
of helical Majorana edge states~\cite{Mo2025topological}, and second-order topological superconducting 
phase with highly tunable MZMs~\cite{Qin2022Topological,Zhang2024TSC,Wang2026SOTSC}. Below we show that this band structure also naturally gives rise
to ideal {\it layer-locked chiral topological superconductivity}.

{\it Layer-locked chiral topological superconductivity.---}Since the BdG Hamiltonian lacks time-reversal symmetry for 
$M_{z}\neq0$, it falls into class D of the tenfold-way classification, where the first-order topological properties are 
characterized by the Chern number~\cite{Schnyder2008Classification,Kitaev2009Periodic,Ryu2010topo}. However, we find that 
the Chern number alone is insufficient to fully capture the 
topological phases and boundary states present in this system. In analogy with insulating systems, a more refined 
quantity---the layer-resolved Chern number---is required. This is defined as~\cite{Li2022Identifying,Li2024High,Han2025Layer} 
\begin{eqnarray}
	C_{z}=\frac{1}{2 \pi}\sum_{E_{n}<0} \int_{BZ} \Omega_{z}^{n}(\bk) d^{2} k, 
\end{eqnarray}
where the subscript $z=\{t,b\}$ indicates the layer label, and $\Omega_{z}^{n}(\bk)$ is the layer-resolved Berry curvature of the $n$th band. 
The explicit form of  $\Omega_{z}^{n}(\bk)$ is given by~\cite{Li2022Identifying,Li2024High,Han2025Layer}  
\begin{eqnarray}
	\Omega_{z}^{n}(\bk)=-2 {\rm Im} \sum_{n^{\prime}\neq n}\frac{\langle u_{n \bk} |P_{z} v_{x} |u_{n^{\prime}\bk} \rangle \langle u_{n^{\prime} \bk} | v_{y} |u_{n \bk} \rangle}{[E_{n^{\prime}}(\bk)-E_{n}(\bk)]^{2}},
\end{eqnarray}
where $|u_{n \bk} \rangle$ denotes the eigenstate, $E_{n}(\bk)$ the corresponding eigenenergy, and $v_{x/y}=\partial_{k_{x}/k_{y}} \mathcal{H}$ the velocity operators. The operator $P_{z}=(1\pm \sigma_z)/2$ projects onto the top $(+)$ and bottom $(-)$ layers, respectively. 
The total Chern number $C$ and the layer-resolved Chern numbers satisfy the natural relation $C=C_{t}+C_{b}$.

It is worth noting that, unlike the total Chern number, the layer-resolved Chern number is not a 
genuine topological invariant. Its values can vary continuously, meaning that exact quantization 
is generally not achievable in a realistic system. Nevertheless, from a practical standpoint, 
near-perfect quantization is effectively equivalent to exact quantization whenever the deviation 
falls below the resolution of a given experimental measurement.

\begin{figure}[t]
	\centering
	\includegraphics[width=0.52\textwidth]{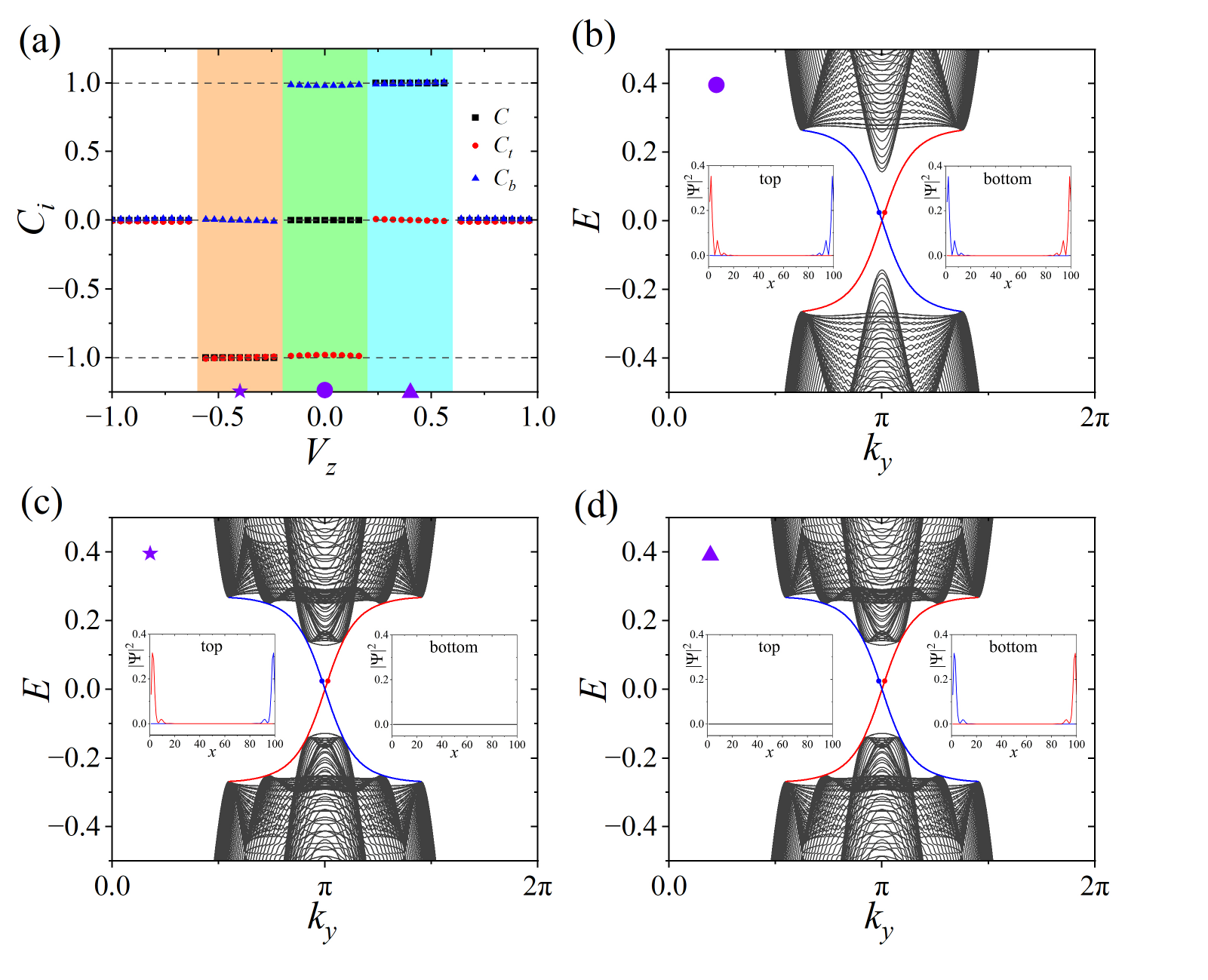}
	\caption{(a) Total Chern number $(C)$ and layer-resolved Chern numbers $(C_{t},C_{b})$ as a function of $V_{z}$. 
The white, yellow, green, and cyan regions correspond to the trivial, top-layer-locking, bilayer-locking, and bottom-layer-locking phases, respectively.  (b)-(d) Energy spectra for a ribbon geometry with open (periodic)
	boundary conditions in the $x (y)$ direction, for $V_{z}=0$, $-0.4$, and $0.4$, respectively.  
The midgap edge-state spectrum is doubly degenerate in (b), whereas no such degeneracy appears in (c) or (d). 
Insets show the layer-resolved probability density profiles of the upper-moving (red) and lower-moving (blue) CMESs on the two $x$-normal edges (with $N_x=100$).  Common parameters are $t=1$, $\mu=3.8$, $\eta=0.5$, $\lambda_{\rm so}=0.5$, $M_{z}=0.5$, and $\Delta=0.3$. 
	}\label{fig2}
\end{figure}

In this system, we find that a nearly quantized layer-resolved Chern number emerges naturally when the Fermi surface is tuned to enclose the 
$\textbf{M}$ point (i.e., $\mu$ close to $4t$). In this regime, the BdG energy gap closes at 
$\textbf{M}$ when $M_{z}=\sqrt{(\mu-4t\pm V_{z})^{2}+\Delta^{2}}$. This gap-closing condition defines the 
phase boundaries separating distinct topological phases. In Fig.~\ref{fig2}(a), we fix 
$(t, \mu, M_{z}, \Delta)=(1,3.8,0.5,0.3)$ and plot the evolution of $(C, C_{t},C_{b})$ as a function 
of  $V_{z}$. The results immediately reveal three distinct topological superconducting phases, 
with phase boundaries located at $V_{z}=\pm0.2$ and $\pm0.6$. Specially, 
for $-0.2<V_{z}<0.2$, we identify a bilayer-locking phase characterized by $C=0$ and $C_{t}=-C_{b}\simeq -1$; for 
$-0.6<V_{z}<-0.2$, a top-layer-locking phase with $C=-1$, $C_{t}\simeq -1$, 
and $C_{b}\simeq0$; and for $0.2<V_{z}<0.6$, a bottom-layer-locking phase with $C=1$, $C_{t}\simeq 0$, 
and $C_{b}\simeq1$.

To further illustrate the layer-locking nature of these topological phases, we compute the spectrum under open boundary conditions along 
$x$ and periodic boundary conditions along $y$, and examine the real-space distribution of the corresponding topological boundary states. In Fig.~\ref{fig2}(b), we show the results at $V=0$---the $\mathcal{PT}$-symmetric limit, corresponding to the bilayer-locking phase. Although the total Chern number vanishes, each edge hosts a pair of counter-propagating Majorana modes, with their spectra crossing at momentum 
$k_{y}=\pi$. The two insets in Fig.~\ref{fig2}(b) show the probability density profile of chiral Majorana modes near zero energy. 
On the same edge (left or right), there is one chiral mode residing on the top layer and another with opposite chirality on the bottom layer. Their degeneracy at $k_{y}=\pi$ signals the absence of hybridization between them, which arises precisely from the mentioned lattice symmetry. This bilayer-locking phase thus realizes a new type of CTSC, characterized by a vanishing total Chern number yet hosting robust layer-locked CMESs.

Under the same boundary conditions, Fig.~\ref{fig2}(c) displays the spectra at $V_{z}=-0.4$. 
The result indicates the presence of a single CMES, consistent with the total Chern number 
$C=-1$. Notably, the chiral Majorana modes at (near) zero energy are (almost) perfectly confined to the top layer (see the inset), in full agreement with the calculated layer-resolved Chern numbers $(C_{t},C_{b})=(-0.999,-0.001)$. Figure ~\ref{fig2}(d) displays the spectra at $V_{z}=0.4$. 
Upon reversing the gate voltage, we obtain $C=1$ and $(C_{t},C_{b})=(0.001,0.999)$. Correspondingly, 
the CMES not only reverses its propagating direction but also becomes confined to the bottom layer (see 
the inset).

The layer-locked behavior observed in Fig.~\ref{fig2} conveys two important messages: (i) a nearly quantized layer-resolved Chern number can serve as a topological invariant, reliably predicting the layer on which the CMESs are confined, even in the presence of strong interlayer coupling; and (ii) the layer-locked chiral topological superconductivity enables fully electric-field tunability [see illustration in 
Fig. \ref{fig1}(b)].

{\it Layer-locked vortex-core MZMs.---}Bulk-vortex correspondence is a unique 
property of CTSC~\cite{Teo2010}. It refers to the relation that the number of robust MZMs ($n_{M}$) in a $\pi$-flux 
vortex is determined by the parity of the Chern number, i.e., $n_{M}=C$ mod 2. This $Z_{2}$ classification 
implies that a $\pi$-flux vortex core hosts a robust MZM when the Chern number is odd, whereas no such mode appears when the Chern number is even (including zero). 

Given the nearly perfect quantization of the layer-resolved Chern number, we expect an analogous 
correspondence to exist between the layer-resolved Chern number and the number of MZMs localized at the vortex core in the corresponding layer. To demonstrate this expectation, 
we simulate a vortex line positioned at the lattice center, as illustrated in Fig.~\ref{fig1}(c). Specifically, we 
impose a real-space dependence on the superconducting order parameter of the form
$\Delta(\br)=\Delta\tanh(\sqrt{x^{2}+y^{2}}/ \xi)e^{i\theta}$, where the phase 
angle $\theta=\arctan(y/x)$ and $\xi$ is the coherence length. Since a single vortex is 
incompatible with periodic boundary conditions, we adopt open boundary conditions along both the $x$ and $y$ directions.

\begin{figure}[t]
	\centering
	\includegraphics[width=0.48\textwidth]{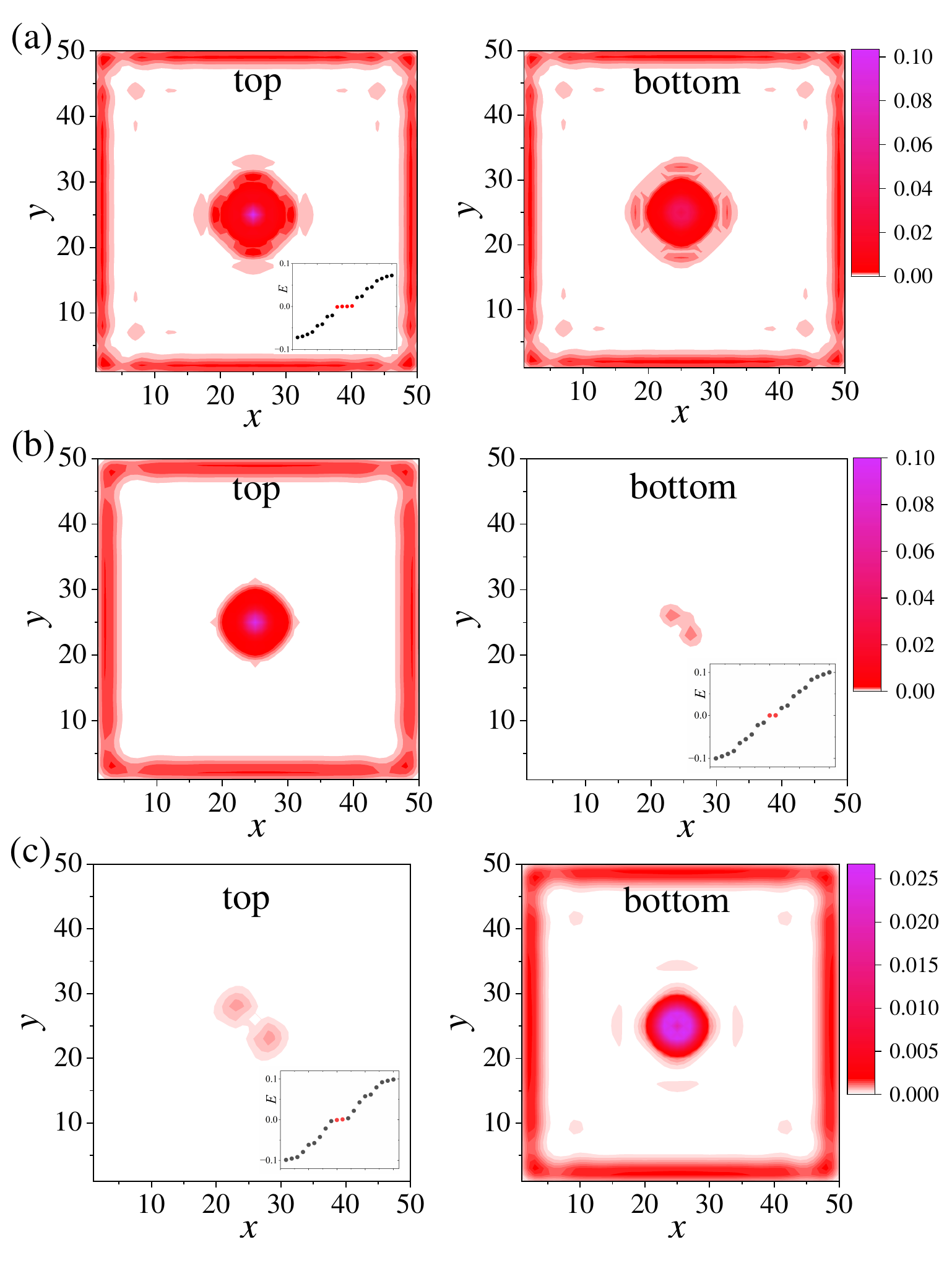}
	\caption{Layer-resolved probability density profiles of MZMs under open boundary conditions in both the $x$ and $y$ directions ($N_{x}=N_{y}=50$), with a single vortex placed at $(x,y)=(25,25)$. Panels (a)–(c) correspond to $V_{z}=0$, $-0.4$, and $0.4$, respectively. Insets show a few eigenenergies closest to $E = 0$. Common parameters are $t=1$, $\mu=3.8$, $\eta=0.5$, $\lambda_{\rm so}=0.5$, $M_{z}=0.5$, $\Delta=0.3$, and $\xi=2$. 
	}\label{fig3}
\end{figure}

Figure~\ref{fig3} shows the probability density profiles of these MZMs in the bilayer-locking [Fig.~\ref{fig3}(a)], top-layer-locking [Fig.~\ref{fig3}(b)], and bottom-layer-locking phases [Fig.~\ref{fig3}(c)], respectively. We note that the eigenvalues of these MZMs are not exactly zero, as our simulations are performed on a finite-size lattice. For the bilayer-locking phase, we observe four MZMs [indicated by the four red dots in the inset of Fig.~\ref{fig3}(a)], with two sharply localized at the vortex core and two extensively distributed along the edges. Focusing on the two vortex-core MZMs situated at the lattice center, we find that they also exhibit the layer-locked feature. This layer-locking property is essential for the coexistence of two robust MZMs at the same vortex. As previously noted, vortex-core MZMs obey a $Z_{2}$ classification. The vanishing total Chern number of this phase suggests that these two MZMs are not topologically protected in the conventional sense. Nevertheless, the nearly quantized layer-resolved Chern number prevents their hybridization, thereby preserving their self-conjugate Majorana nature. 
The presence of twofold Majorana zero modes at the vortex core in a similar phase has also been 
noted in an earlier study~\cite{Zhang2024TSC}, with symmetry protection invoked to explain their lack of hybridization.

The layer-locked behavior of the vortex-core MZMs becomes even more evident in the other two single-layer-locking phases, where 
$|C|=1$. In these phases, the total Chern number ensures the presence of one robust MZM at the vortex core and one at the edge. For the top-layer-locking phase, we observe that the vortex-core MZM is strongly confined to the top layer, as reflected by the pronounced weight difference between the top and bottom layers (top/bottom $\sim56$) in Fig.~\ref{fig3}(b). Upon reversing the gate voltage to drive the system into the bottom-layer-locking phase, the vortex-core MZM is correspondingly transferred to the bottom layer, 
as shown in Fig.~\ref{fig3}(c).

{\it Layer-resolved thermal Hall effects.---}Thermal Hall effect describes the generation of a transverse current in response to a longitudinal temperature gradient~\cite{Qin2011Energy,Sumiyoshi2013Quantum,Wang2021Berry,Mo2025Coexistence}, as illustrated in Fig. \ref{fig1}(d).  In the low-temperature limit ($T\rightarrow0$), 
the thermal Hall coefficient in a CTSC takes the quantized form $\kappa^{xy}=C\kappa_{0}$~\cite{Read2000Paired}, where 
$\kappa_{0}=\pi k_{B}^{2}T/12\hbar$, $k_{B}$ is the Boltzmann constant (hereafter set to unity), 
and $T$ is the temperature. By analogy with the definition of the layer-resolved Hall conductivity, 
we define the layer-resolved thermal Hall coefficient as 
\begin{eqnarray}
	\kappa_{z}^{xy}=\frac{1}{2\hbar T} \int_{-\infty}^{+ \infty} E^{2} \sigma_{z}(E) f^{\prime}(E) dE,\label{thermalH}
\end{eqnarray}
where $f(E)=1/\exp[(E/T)+1]$ is the Fermi-Dirac distribution function, and the layer-resolved quantity $\sigma_{z}(E)$ takes the form 
\begin{eqnarray}
	\sigma_{z}(E)=\sum_{n}\int_{E_{n}(\bk)<E} \Omega_{z}^{n}(\bk) \frac{d^{2} k}{(2\pi)^2}.\label{sigmaz}
\end{eqnarray}
Evidently, from the relation $\Omega^{n}=\Omega_{t}^{n}+\Omega_{b}^{n}$, it follows that the total thermal Hall coefficient satisfies $\kappa^{xy}=\kappa_{t}^{xy}+\kappa_{b}^{xy}$.

\begin{figure}[t]
	\centering
	\includegraphics[width=0.5\textwidth]{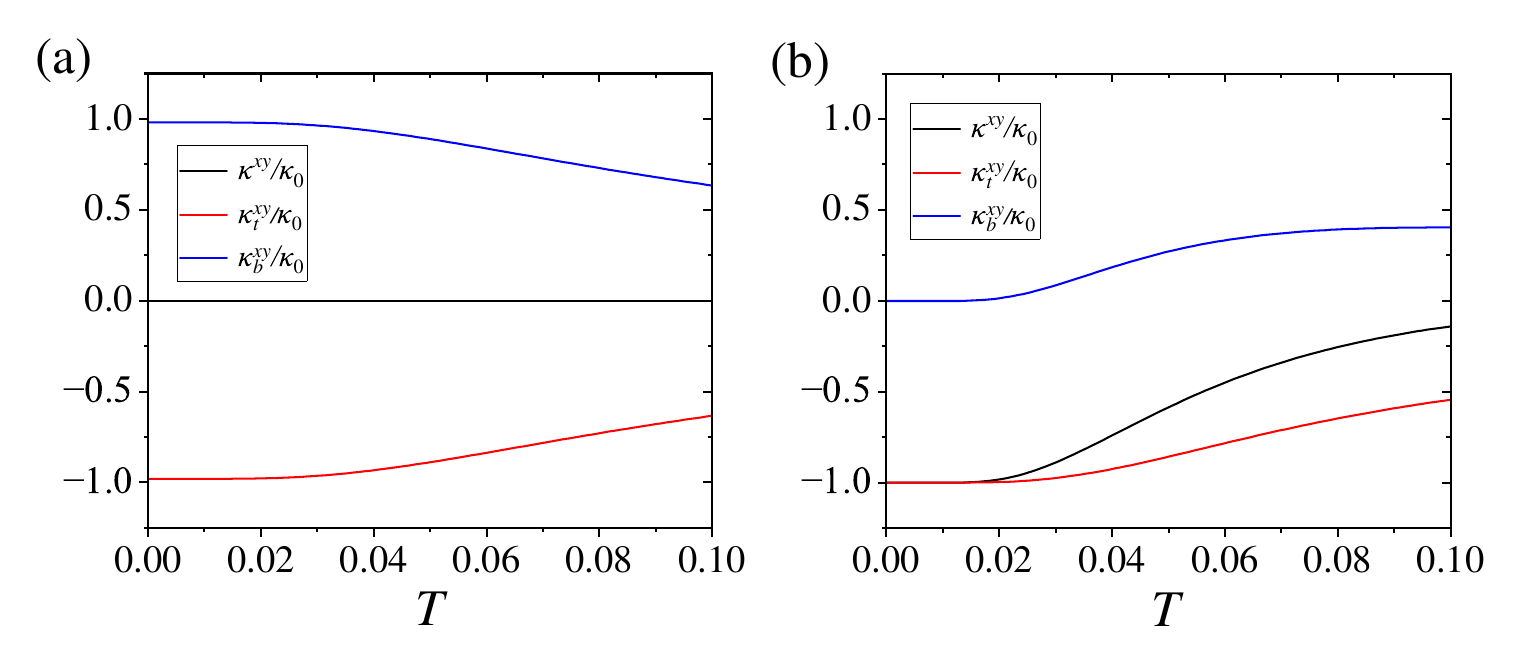}
	\caption{Layer-resolved thermal Hall coefficients as functions of temperature. The solid black, red, and blue lines represent the total, top-layer, and bottom-layer thermal Hall coefficients, respectively. Panels (a) and (b) correspond to $V_{z}=0$ and $-0.4$, respectively. Common parameters are: $t=1$, $\mu=3.8$, $\eta=0.5$, $\lambda_{\rm so}=0.5$, $M_{z}=0.5$, and $\Delta=0.3$. 
	}\label{fig4}
\end{figure}

Figures~\ref{fig4}(a) and~\ref{fig4}(b) show the temperature evolution of the thermal Hall coefficients in the bilayer-locking and top-layer-locking phases, respectively. Similar to the cases of CMESs and vortex-core MZMs, the thermal Hall effect in the low-temperature regime also exhibits layer-locked behavior. In the bilayer-locking phase, we find that as $T\rightarrow0$, $\kappa^{xy}=0$ and $\kappa_{t}^{xy}=-\kappa_{b}^{xy}= -0.981\kappa_{0}$. For the parameters chosen here, we have $(C,C_{t},C_{b})=(0,-0.981,0.981)$. It is evident that the relation $\kappa_{z}^{xy}=C_{z}\kappa_{0}$ holds exactly. In the top-layer-locking phase, as $T\rightarrow0$, we obtain $\kappa^{xy}=-\kappa_0$, $\kappa_{t}^{xy}=-0.999\kappa_{0}$, and $\kappa_{b}^{xy}=-0.001\kappa_{0}$, again satisfying the relation
$\kappa_{z}^{xy}=C_{z}\kappa_{0}$. In this phase, the total thermal Hall coefficient is dominated by the top layer 
at low temperatures. However, as the temperature rises, bulk excitations from the bottom layer begin to contribute, 
causing the total thermal Hall coefficient to gradually deviate from that of the top layer. 
Reversing the gate voltage to enter the bottom-layer-locking phase would switch the dominant layer and 
reverse the direction of the transverse current, as illustrated in Fig.~\ref{fig1}(d).

{\it Discussions and conclusions.---}We have identified a universal mechanism for realizing 
layer-locked topological phases. On this basis, we propose the concept of layer-locked chiral topological 
superconductivity as the superconducting analogue of the quantum anomalous layer Hall effect. 
Using a bilayer model, we demonstrate that nearly quantized layer-resolved Chern numbers---alongside 
a variety of layer-locked topological phenomena---can be achieved simply by tuning the topological phase 
transition to occur at a momentum on the Brillouin zone boundary, where the interlayer coupling effectively vanishes.
Our study admits two natural generalizations. First, the minimal bilayer setup can be 
straightforwardly extended to multilayered systems, which would enable even richer layer-locked physics, 
including larger Chern numbers and the selective design of the number and chirality of chiral edge states 
across different layers. Second, the layer-locked topological phases can be generalized either from chiral to helical phases 
or from first-order to 
second-order topological phases~\cite{Li2024LHE}. Our findings not only broaden the
landscape of layer-locked phenomena but also provides a new
pathway for realizing highly tunable CTSCs, potentially offering new platforms for topological 
quantum computation and quantum thermoelectric devices.

Experimentally, scanning tunneling microscopy offers a direct probe of layer-locked CMESs and 
vortex-core MZMs. The layer thermal Hall 
effects can be detected in analogy with measurements of the layer Hall effect—specifically, 
by observing the sign change in the thermal Hall conductivity upon reversing the gate voltage. 
For experimental realization of our proposal, layered superconductors with nonsymmorphic symmetries 
represent ideal candidate systems. A particularly promising example is the iron-based superconductor 
FeSe, whose lattice structure and underlying symmetries match those of our bilayer setup~\cite{Young2017Filling}. Although 
antiferromagnetic order may not coexist with superconductivity in this material, the magnetic exchange 
field can be effectively replaced by a Zeeman field generated through an external magnetic field.

{\it Acknowledgements.---}This work is supported by the Fundamental and Interdisciplinary Disciplines Breakthrough Plan of the Ministry of Education of China (Grant No. JYB2025XDXM403) and the Guangdong Basic and Applied Basic Research Foundation (Grant No. 2023B1515040023).

\bibliography{layerH}

\end{document}